\documentclass[twocolumn,showpacs,preprintnumbers,amsmath,amssymb]{revtex4}
\input psfig.sty

\usepackage{graphicx}
\usepackage{dcolumn}
\usepackage{bm}
\usepackage{epsfig}

\newcommand{\om}{\Omega_M}
\newcommand{\ok}{\Omega_k}
\newcommand{\ox}{\Omega_x}
\newcommand{\oll}{\Omega_\Lambda}

\begin{document}


\title{ Constraining $H_0$ in General Dark Energy Models \\from Sunyaev-Zeldovich/X-ray Technique and Complementary Probes}

\author{R. F. L. Holanda$^1$}\email{holanda@astro.iag.usp.br}
\author{J. V. Cunha$^1$}\email{cunhajv@astro.iag.usp.br}
\author{L. Marassi$^{2}$}\email{luciomarassi@ect.ufrn.br}
\author{J. A. S. Lima$^{1}$}\email{limajas@astro.iag.usp.br}
\vspace{0.5cm}

\affiliation{$^1$Departamento de Astronomia (IAGUSP),  Universidade
de S\~{a}o Paulo \\ Rua do Mat\~ao, 1226 -- 05508-900, S\~ao Paulo,
SP, Brazil} \affiliation{$^{2}$Escola de Ci\^encia e
Tecnologia,UFRN, 59072-970, Natal, RN, Brasil}

\date{\today}

\begin{abstract}
In accelerating dark energy models, the estimates of $H_0$ from
Sunyaev-Zel'dovich effect (SZE) and X-ray surface brightness of
galaxy clusters may depend  on the  matter content ($\Omega_{M}$),
the curvature ($\Omega_K$) and the equation of state parameter
($\omega$). In this article,  by using a sample of 25 angular
diameter distances from galaxy clusters obtained through SZE/X-ray
technique, we constrain $H_0$ in the framework of  a general
$\Lambda$CDM models (free curvature) and a flat XCDM model with
equation  of state parameter $\omega=p_x/\rho_x$
($\omega$=constant). In order to broke the degeneracy on the
cosmological parameters, we apply a joint analysis involving the
baryon acoustic oscillations (BAO)  and the CMB \textit{Shift
Parameter} signature. By neglecting systematic uncertainties, for
nonflat $\Lambda$CDM cosmologies we obtain $H_0=73.2^{+4.3}_{-3.7}$
km s$^{-1}$ Mpc$^{-1}$(1$\sigma$) whereas for a flat universe with
constant equation of state parameter we find
$H_0=71.4^{+4.4}_{-3.4}$ km s$^{-1}$ Mpc$^{-1}$(1$\sigma$). Such
results are also in  good agreement with independent studies from
the {\it{Hubble Space Telescope}} key project and recent estimates
based on {\it Wilkinson Microwave Anisotropy Probe}, thereby
suggesting that the combination of these three independent phenomena
provides an interesting method to constrain the Hubble constant. In
particular, comparing these results with a recent determination for
a flat $\Lambda$CDM model using only the SZE technique and BAO
[Cunha et al. MNRAS {\bf 379}, L1 2007],  we see that the geometry
has a very weak influence on $H_0$ estimates for this combination of
data.

\end{abstract}

\maketitle

\section{INTRODUCTION}
\label{sec:intro}

Clusters of galaxies are the largest gravitationally bound
structures in the universe and they can be regarded as being
representative of the universe as a whole. An important phenomena
occurring in clusters is the Sunyaev-Zel'dovich effect (SZE), a
small distortion of the Cosmic Microwave Background (CMB) spectrum,
provoked by the inverse Compton scattering of the CMB photons
passing through a population of hot electrons
\cite{SunZel72,sunzel80,itoh,Bartlett1,Bartlett2,Kohyama09,Lavalle10},
The SZE is quite independent of the redshift, so it provides an
useful tool for studies of intermediate and high redshift galaxy
clusters, where the cosmological model adopted plays an important
role (for a reviews see \cite{Birki99,Carlstrom02}).

When the X-ray emission of the intracluster medium (ICM) is combined
with the SZE, it is possible to estimate the angular diameter
distance (ADD) ${\cal{D}}_A$. In other words, the SZE/X-ray method
provides distances to the clusters and consequently a measure of the
Hubble parameter $H_0$. The main advantage of this method for
estimating $H_0$ is that it does not rely on the extragalactic
distance ladder, being fully independent of any local calibrator
\cite{Carlstrom02,Jones05}.

It should be stressed that the determination of Hubble parameter has
a practical and theoretical importance to many astrophysical
properties and cosmological observations \cite{Freedman00}. Komatsu
{\it{et al.}} have shown that CMB studies can not supply strong
constraints to the value of $H_0$ on their own \cite{komatsu}. This
problem occurs due to the degeneracy on the parameter space and may
be circumvented only by using independent measurements of $H_0$
\cite{Hu05}.

In this connection, Sandage and collaborators \cite{Sandage06}
announced the results from their HST programme, $H_0=62\pm 5$
km/s/Mpc, whereas Van Leeuwen {\it{et al.}} \cite{VanLeeuwen07}
revised Hipparcos parallaxes for Cepheid distance and obtained
higher values than previous results advocated by  Sandage {\it et
al.} \cite{Sandage06} and Freedman {\it et al.} \cite{WFred01}
groups.

Later on, Riess {\it{et al.}} \cite{riess2009} reported results from
a program to determine the Hubble parameter to $\approx 5$\%
precision from a refurbished distance ladder based on extensive use
of differential measurements. They obtained $H_{0} =74.2 \pm 3.6$
km/s/Mpc, a $4.8$\% uncertainty including both statistical and
systematic errors. Indeed, some estimates of $H_0$ have yielded
$\simeq 74$ km/s/Mpc \cite{komatsu,WFred01,riess2009}.

More recently, by studying time delay from gravitational lens, two
groups obtained $H_0$ estimates in a flat $\Lambda$CDM framework,
Fadely {\it{et al.}} \cite{FalelySuyu10} adding constraints from
stellar population synthesis models obtained $H_0=
79.3^{+6.7}_{-8.5}$ km/s/Mpc ($1\sigma$, without systematic errors),
and Suyu {\it{et al.}} \cite{suyu}, in combination with WMAP
obtained $H_0= 69.7^{+4.9}_{-5.0}$ km/s/Mpc ($1\sigma$, without
systematic errors). The importance to access the distance scale by
different methods, and, more important, in a manner independent of
any distance calibrator has been discussed  in the review paper by
Jackson \cite{Jackson07}.

A couple of years ago,  Cunha, Marassi and Lima \cite{Cunha07}
(henceforth CML), derived new constraints on the matter density and
Hubble parameters ($\Omega_m, H_0$),  by using the SZE/X-ray
technique in the framework of a flat $\Lambda$CDM model. By
considering a sample of 25 galaxy clusters compiled by De Filippis
{\it{et al.}} (2005) \cite{DeFilipp05}, the degeneracy on the
$\Omega_{m}$ parameter was broken trough a joint analysis combining
the SZE/X-ray data with the recent measurements of the baryon
acoustic oscillation (BAO) signature from SDSS catalog
\cite{Eisenstein05,Percival10}. The main advantage of the method is
that we do not need to adopt a fixed cosmological concordance model
in our analysis, as usually done in the literature
\cite{Kobayashi96,carlstrom2002,Mas01,Reese02,Reese04,Jones05,schmidt,Boname06,Itoh9806}.
For a flat $\Lambda$CDM our joint analysis yielded $\Omega_m=
0.27^{+0.03}_{-0.02}$ and $H_0= 73.8^{+4.2}_{-3.3}$ km/s/Mpc
($1\sigma$, neglecting systematic uncertainties).

On the other hand, astronomical observations in the last decade have
suggested that our world behaves like a spatially flat scenario,
dominated by cold dark matter (CDM) plus an exotic component endowed
with large negative pressure, usually named dark energy
\cite{perm98,Riess,Astier06,bar10}. In the framework of general
relativity, besides the cosmological constant, there are several
candidates for dark energy, among them: a vacuum decaying energy
density, or a time varying $\Lambda(t)$ \cite{Lambdat}, the
so-called ``X-matter" \cite{xmatt}, a relic scalar field
\cite{scalarf}, a Chaplygin gas \cite{Chap1,Chap2}, and cosmologies
proposed to reduce the dark sector, among them,  models with
creation of cold dark matter particles \cite{CCDM}.  For a scalar
field component and ``X-matter" scenarios, the equation of state
parameter may be a function of the redshift (see, for example,
\cite{Efsta99}) or still, as has been discussed by many authors,  it
may violate the null energy condition \cite{Phantom}.

In this work, we relax the flat geometry condition of the cosmic concordance model ($\Lambda$CDM).
This procedure will prove the robustness of the
previous $H_{0}$ estimate using the SZE/X-ray technique. In
addition, in order to test the real dependence of the method with
the equation of state parameter, we compare the predictions of the
general $\Lambda$CDM model and ``X-matter" cosmologies \cite{xmatt}.
For the ``X-matter" model we assume a flat XCDM cosmology with
constant equation of state (EoS) parameter. We use the 25
ellipsoidal clusters from De Filippis {\it{et al.}} (2005)
\cite{DeFilipp05}. To broke the degeneracy on the basic cosmological
parameters, we apply a joint analysis using BAO
\cite{Eisenstein05,Percival10} and the CMB probe known as shift
parameter \cite{Bond:1997wr,Nesseris:2006er,Davis,Elgaroy}.

The paper is organized as follows. In section II, we present the
basic equations to angular distance and models studied. In section
III, we give a short description of the observational data we have
used. The corresponding constraints on the cosmological parameters
are investigated in section IV. The article is ended with a summary
of the the main results in the conclusion section.

\section{Basic Equations and Models}

Let us now assume that the Universe is well described by an
homogeneous and isotropic geometry
\begin{equation}
\label{line_elem}
  ds^2 = dt^2 - a^{2}(t) \left(\frac{dr^2}{1-k r^2} + r^2 d\theta^2+
      r^2{\rm sin}^{2}\theta d \phi^2\right),
\end{equation}
where $a(t)$ is the scale factor and $k= 0, \pm 1$ is the curvature
parameter. Throughout we use units such that $c=1$.

In such a background, the angular diameter distance ${\cal{D}}_A$
reads \cite{Lima03,LA00,DeFilipp05}
\begin{equation}
{\cal{D}}_A = \frac{3000h^{-1}}{(1 +
z)\sqrt{|\ok|}}S_k\left[\sqrt{|\ok|}\int_{0}^{z}\frac{dz'}{E(z')}\right]
\,\mbox{Mpc}, \label{eq1}
\end{equation}
where $h=H_0/100$ km s$^{-1}$ Mpc$^{-1}$ (henceforth we use this
notation for our  Hubble parameter estimates), the function
$E(z)={H(z)/H_{0}}$ is the normalized Hubble parameter which defined
by the specific cosmology adopted, $\ok$ is the curvature parameter,
and $S_k(x)=\sin{x}, x$, $\sinh{x}$ for $k=+1, 0$, $-1$,
respectively.

In this paper, we consider that the Universe is driven by cold dark
matter ($\Omega_M$) plus a dark energy component ($\Omega_x$), with
constant EoS parameter ($\omega \equiv {p}/{\rho}$). Below we will
summarize the function $E(z)$ of the cosmological models adopted in
this paper.

(i) $\Lambda$CDM model.  By allowing deviations from flatness, the
normalized Hubble parameter is given by
\begin{equation}E^2(z)={\om(1+z)^3}+{\oll}+{\ok (1+z)^2},\label{eq4}
\end{equation}
where $\ok=1-\oll - \om$.

(ii) Parametric Dark Energy model (XCDM). In this case, the
normalized Hubble parameter reads:

\begin{equation} E^2(z)={\om  (1+z)^3}+{\ox (1+z)^{3(1+w)}}+{\ok (1+z)^2}. \label{eq5}
\end{equation}
where $\ok=1-\ox - \om$. For $\omega=-1$, the above XCDM expression
reduces to the previous $\Lambda$CDM case ($\ox \equiv \oll$). When
the EoS parameter  of dark energy is restricted on the interval
$-1\leq\omega<0$, it may be represented by a scalar field (quintessence), and whether $\omega <-1$ thereby violating
the null energy condition is the case of phantom dark energy
\cite{Phantom}.

Given the above expressions, we see clearly that ${\cal{D}}_A$ is a
function of $z$, $h$ and depending on the model adopted of $\omega$,
$\Omega_M$ and $\Omega_x$. Due to the excessive number of free  parameters in
XCDM models, in our analysis of this dark energy model we will concentrate our attention to the flat case.

\section{Theoretical Method and Samples}

It should be recalled that the basic aim here is to constrain the
Hubble and other cosmological parameters of the above
models. The method adopted is primarily based on the angular
diameter distances, ${\cal{D}}_A$, furnished by the SZE/X-ray
technique. The degeneracies on the remaining cosmological parameters
will be broken by a joint analysis involving  BAO and CMB signature
(shift parameter).

\subsection{SZE/X-ray}

By using an elliptical 2-Dimensional $\beta$-model to describe the
galaxy clusters geometry, De Filippis and coworkers
\cite{DeFilipp05} derived  ${\cal{D}}_A$ measurements for 25
clusters from two previous compilations: one set of data compiled by
Reese {\it{et al.}} \cite{Reese02}, composed by a selection of 18
galaxy clusters distributed over the redshift interval $0.14 < z <
0.8$, and the sample of Mason {\it{et al.}} \cite{Mas01}, which has
7 clusters from the X-ray limited flux sample of Ebeling {\it{et
al.}} (1996) \cite{Ebeling96}. These two previous compilations used
a spherical isothermal $\beta$ model to describe the clusters
geometry.

In the CML paper \cite{Cunha07}, the Abell 773 cluster was excluded
from the statistical analysis due to its large contribution to the
$\chi^2$.  Now, in order to preserve the completeness of the
original data set, these statistical cut-offs arguments will be
neglected, and, as such, all the 25 clusters from the original De
Fillipis {\it{et al.}} \cite{DeFilipp05} sample will be considered.

\subsection{Baryon Acoustic Oscillations (BAO)}

The detection of a peak in the large-scale correlation function
of luminous red galaxies selected from the Sloan Digital Sky Survey (SDSS) Main
Sample gave rise to a powerful cosmological probe, often referred to as BAO scale.

Basically, the peak detected at the scale  of 100 $h^{-1}$ Mpc,
happens due to the baryon acoustic oscillations in the primordial
baryon-photon plasma prior to recombination, and, such a result,
provides a suitable ``standard ruler'' for constraining dark energy
models \cite{Eisenstein05,Percival10}.

The relevant distance measure is the so-called dilation
scale  that can be modeled as the cube root of the radial dilation
times the square of the transverse dilation, at the typical redshift
of the galaxy sample, $z=0.35$ \cite{Eisenstein05}:
\begin{equation}
  D_V(z)=[D_A(z)^2 z/H(z)]^{1/3}\ ,
\end{equation}

Recalling that the comoving size of the sound horizon at $z_{ls}$ is
$\sim 1/\sqrt{\om H_0^2}$, it was pointed out that the combination
$A(z)=D_V(z)\sqrt{\om H_0^2}/z$ is independent of $H_0$, and this
dimensionless combination is also well constrained by the BAO data
\begin{equation}
A(0.35)=D_V(0.35)\frac{\sqrt{\om H_0^2}}{0.35}=0.469\pm 0.017.
\end{equation}

The  BAO quantity $A(0.35)$ is exactly what we add to the
$\chi^{2}$, in the joint statistical analysis of the cosmological
models studied here.

\subsection{CMB - Shift Parameter}

Another interesting cosmological probe to dark energy models is the
so-called CMB shift parameter. Such a quantity is encoded in the
location $l_1^{TT}$ of the first peak of the angular (CMB) power
spectrum \cite{Bond:1997wr,Nesseris:2006er}
\begin{equation}
\theta_A \equiv \frac{r_s(z_{ls})}{D_A(z_{ls})}\ ,
\end{equation}
where $z_{ls}$ is the redshift of the last scattering surface and
$D_A(z_{ls})$ is the angular distance to the last-scattering
surface. The quantity, $r_s(z_{ls})\sim 1/\sqrt{\om H_0^2}$, is the
comoving size of the sound horizon at $z_{ls}$.

By using the {\it WMAP} three years result \cite{komatsu} Davis
{\it{et al.}} \cite{Davis} converted the location of the first peak
in a reduced distance to the last-scattering surface
\begin{equation} R=\sqrt{\frac{\om}{|\ok|}}\;
 S_k\left[\sqrt{|\ok|}\int_0^{z_{\rm ls}}\frac{dz}{E(z)}\right]=1.71\pm
 0.03.
\end{equation}

The robustness of the shift parameter has been tested by Elgaroy \&
Multam{\"a}ki \cite{Elgaroy}  and compared to fits of the full CMB
power spectrum. As a result, it is now widely believed that the
shift parameter is an accurate geometric measure even for
non-standard cosmologies. It is weakly dependent on $h$, since
$z_{ls}=z_{ls}(h,\Omega_{M},\Omega_{b})$ is a smooth function of
$h$, and the degeneracies that arise from using R rather than
fitting the full CMB power spectrum are well constrained by other
data such as BAO and SZE/X-ray.

In the following computations for the theoretical shift parameter we
will apply the correction for $z_{ls}$ suggested in Refs.
\cite{Elgaroy,sze1}.  In addition, independent of the adopted dark
energy model, we keep the value $R=1.70\pm0.03$ for a flat universe
and $R=1.71\pm0.03$ for nonzero cosmic curvature \cite{sze1}.

\begin{figure*}\label{Fig2}
\centerline{
\epsfig{figure=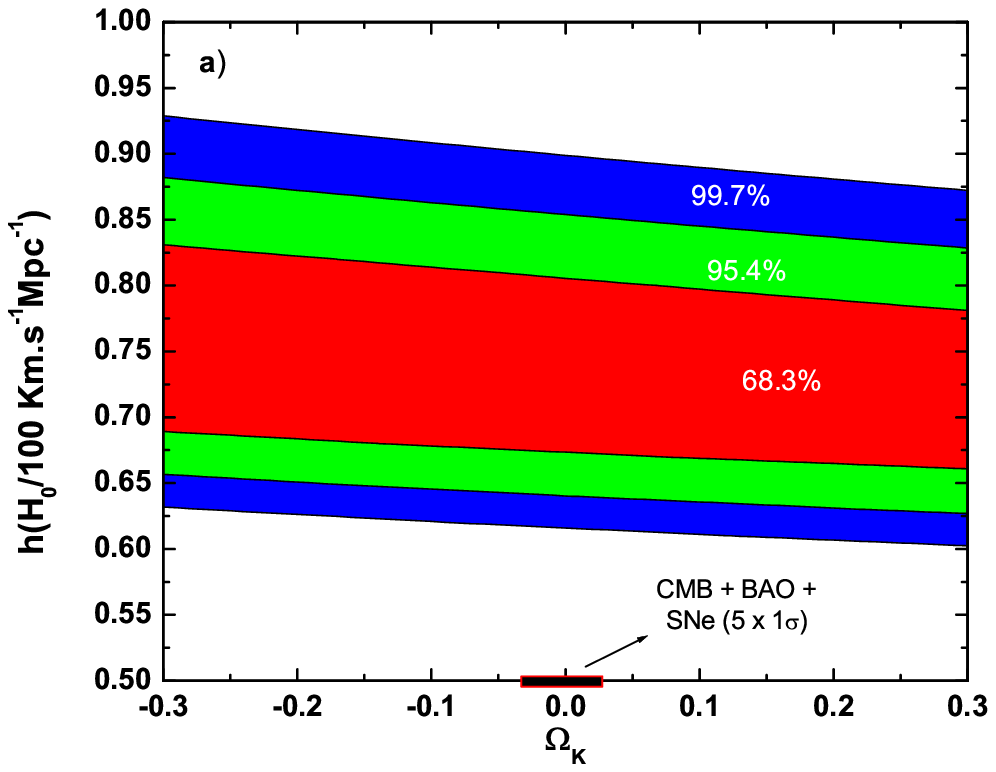,width=2.3truein,height=2.1truein}
\epsfig{figure=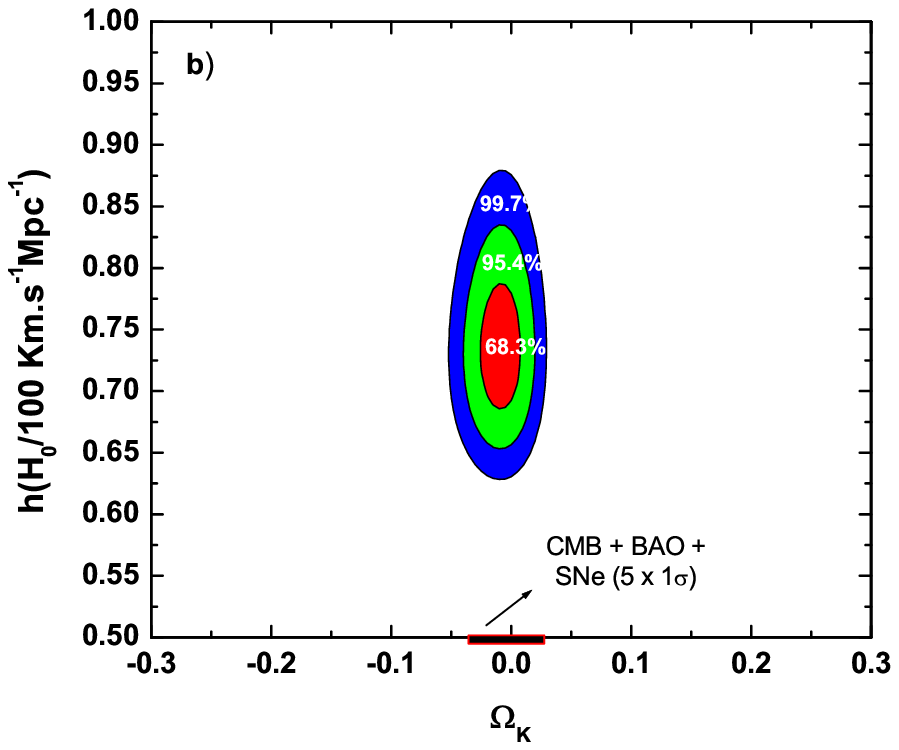,width=2.3truein,height=2.1truein}
\epsfig{figure=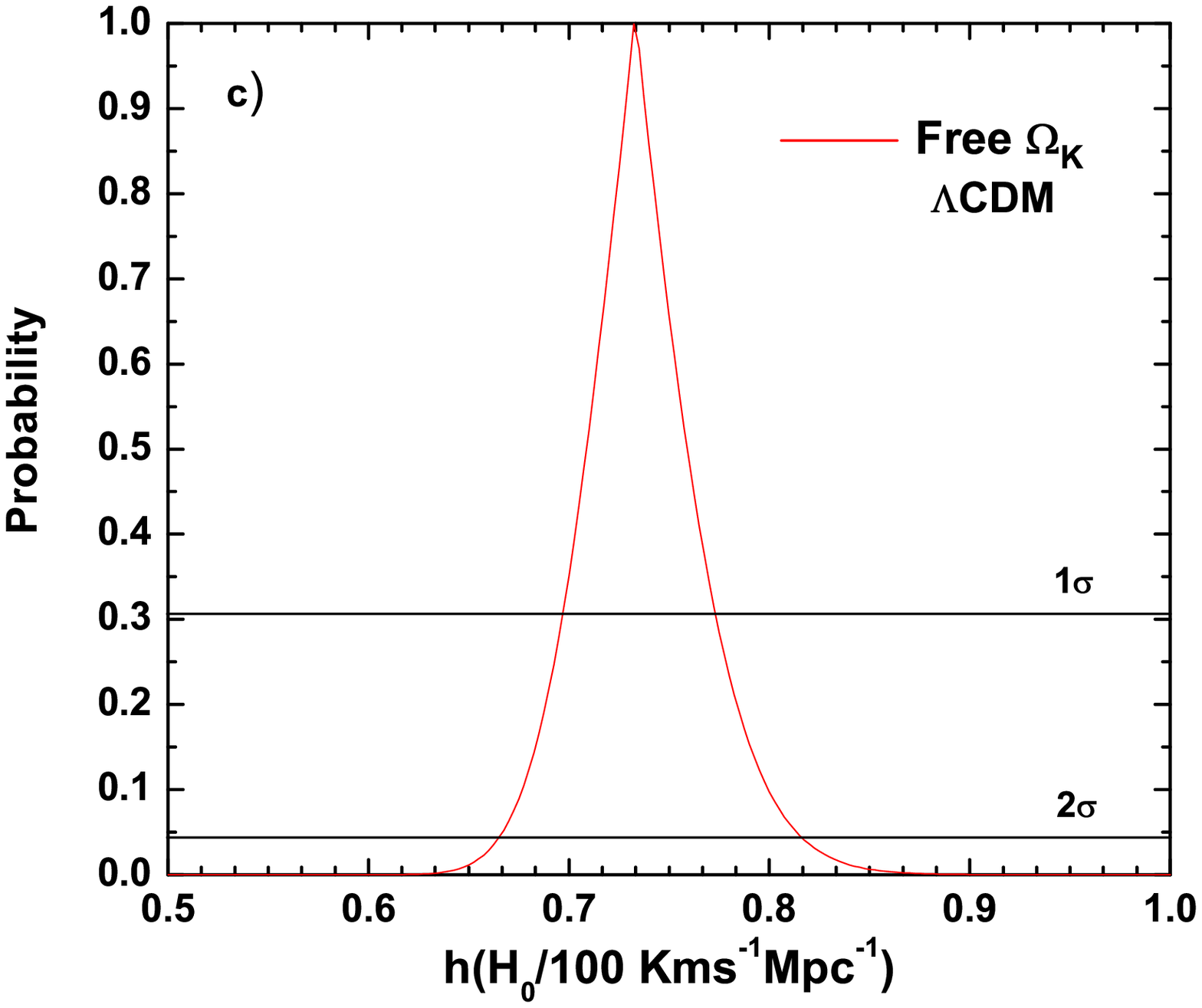,width=2.3truein,height=2.1truein} \hskip
0.1in} \caption{$\Lambda$CDM Model. {\bf a)} Contourns in the $(
h,\Omega_K)$ plane using only SZE/X-ray technique. Note that the
constraints on $h$ are very weekly dependent on the curvature of
Universe. This result has been derived by marginalizing over
$\Omega_{M}$. {\bf b)} Combining different probes. Contours at
$68.3$\%, $95.4$\% and $99.7$\% c.l. in the $( h,\Omega_{K})$ plane
trough a joint analysis involving SZE/X-ray + BAO + Shift Parameter.
{\bf c)} Probability of the $h$ parameter. The horizontal lines
represent cuts of $68.3$\% and $95.4$\%of statistical confidence.
The best-fit result is $h=0.73$ with a reduced $\chi^2_{red} =
1.12$.}
\end{figure*}

\section{Analysis and Results}

In our statistical procedure we apply a maximum likelihood analysis
determined by a $\chi^2$ statistics
\begin{equation}
\chi^2(z|\mathbf{p}) = \sum_i { ({\cal{D}}_A(z_i; \mathbf{p})-
{\cal{D}}_{Ao,i})^2 \over \sigma_{{\cal{D}}_{Ao,i}}^2},
\end{equation}
where ${\cal{D}}_{Ao,i}$ is the observational angular diameter
distance, $\sigma_{{\cal{D}}_{Ao,i}}$ is the uncertainty in the
individual distance and  $\mathbf{p}$ is the complete set of
parameters. For the $\Lambda$CDM model $\mathbf{p} \equiv (h, \om,
\Omega_{\Lambda})$  and $\mathbf{p} \equiv (h, \omega, \om)$  for
the flat XCDM model.


 All the systematic effects still need to be considered. The common errors
are: SZ calibration $\pm 8$\%, X-ray flux calibration $\pm 5$\%,
radio halos $+3$\%, and X-ray temperature calibration $\pm 7.5$\%.


\subsection{$\Lambda$CDM Model}

Allowing for deviations from flatness, we first consider the general
$\Lambda$CDM model as described by Eq. (\ref{eq4}).

In Figure 1a, we display the contours ($68.3$\%, $95.4$\% and
$99.7$\% c.l.) in the $( h, \Omega_{k})$ plane  provided by the
diameter angular distances from SZE/X-ray technique (we have
marginalized over $\Omega_M$). We see clearly a degeneracy between
$\Omega_{k}$ e $h$, and, therefore, the possible values for $h$ are
very weekly constrained from SZE/X-ray alone. Note that the  h
parameter lies on the interval  $0.63<h<0.93$ at 99.7\% c.l. (1 free
parameter).

In Figure 1b, we show the results for a  joint analysis involving
SZE/X-ray + BAO + Shift Parameter. In this case we obtain $h=
0.733^{+0.042}_{-0.037}$,
 $\ok=-0.010^{+0.012}_{-0.013}$ and $\chi^2_{min}=28.12$ at
$68.3$\% (c.l.). Its reduced value, that is, taking into account the
associated degrees of freedom is $\chi^2_{red}=1.12$ thereby showing
that the fit is very good.

In Figure 1c, we display the likelihood function of the $h$
parameter by using the SZE/X-ray data + BAO + Shift Parameter. To
obtain this graph we have marginalized over $\Omega_M$ and
$\Omega_{\Lambda}$ parameters. The horizontal lines are cuts in the
probability regions of $68.3$\% and $95.4$\%. This plot is very
similar to the Fig. (4) of the previous CML paper \cite{Cunha07} for
a flat $\Lambda$CDM model. Therefore, it is safe to conclude that
the constraints on $h$ derived here are independent from the
geometry of the Universe.



\subsection{Flat XCDM Model}

As we know, some dark energy candidates are phenomenologically
described by an equation of state of the form, ${p}=\omega {\rho}$,
where $\omega$ is a constant parameter. In the flat case, the
normalized Hubble parameter is readily obtained by taking
$\Omega_k=0$ in Eq. (\ref{eq5}). In this context, by relax the usual
imposition $\omega\geq -1$ we investigate some implications for the
so-called phantom dark energy. The basic idea here is to test the
sensibility of the SZE/X-ray data + BAO + Shift Parameter with
respect to the $\omega$ parameter.  In addition, we also study the
influence of XCDM on the Hubble constant determination, thereby
performing also a direct comparison to the $\Lambda$CDM model
($\omega=-1$).

In Figure 2a, we display the $( h,\omega)$ plane (marginalizing over
$\Omega_{M}$) using the 25 angular diameter distances from galaxy
clusters. The limits on $h$ in the $( h,\omega)$ plane is wider than
in the $\Lambda$CDM models, but the $h$ value has a very weak
dependence on $\omega$ parameter. We stress that we have explored a
large range of this parameter ( $-3<\omega<0.5$).

In Figure 2b, we show the results when a joint analysis involving
the SZE/X-ray data set + BAO + Shift Parameter is performed. We find
$h= 0.714^{+0.044}_{-0.034}$, $\omega=-0.76^{+0.19}_{-0.28}$ and
$\chi^2_{min}=28.35$ at $68.3$\% confidence level, whereas its
reduced value is $\chi^2_{red}=1.13$. Again, we have an excellent
fit.
\begin{figure*}\label{Fig4}
\centerline{
\epsfig{figure=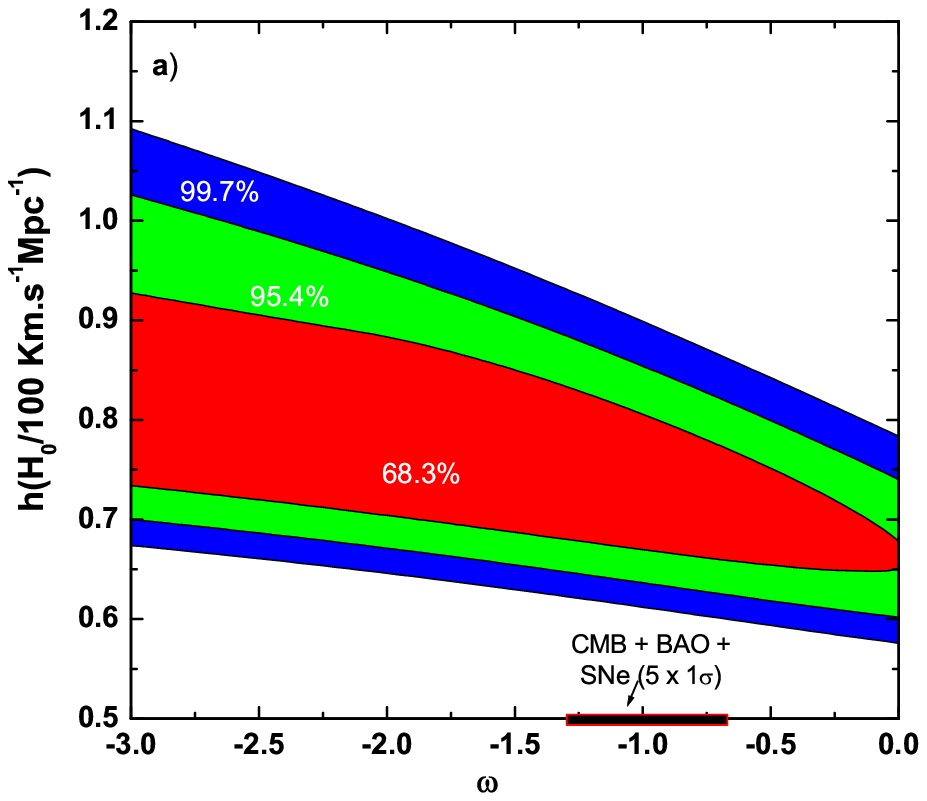,width=2.3truein,height=2.1truein}
\epsfig{figure=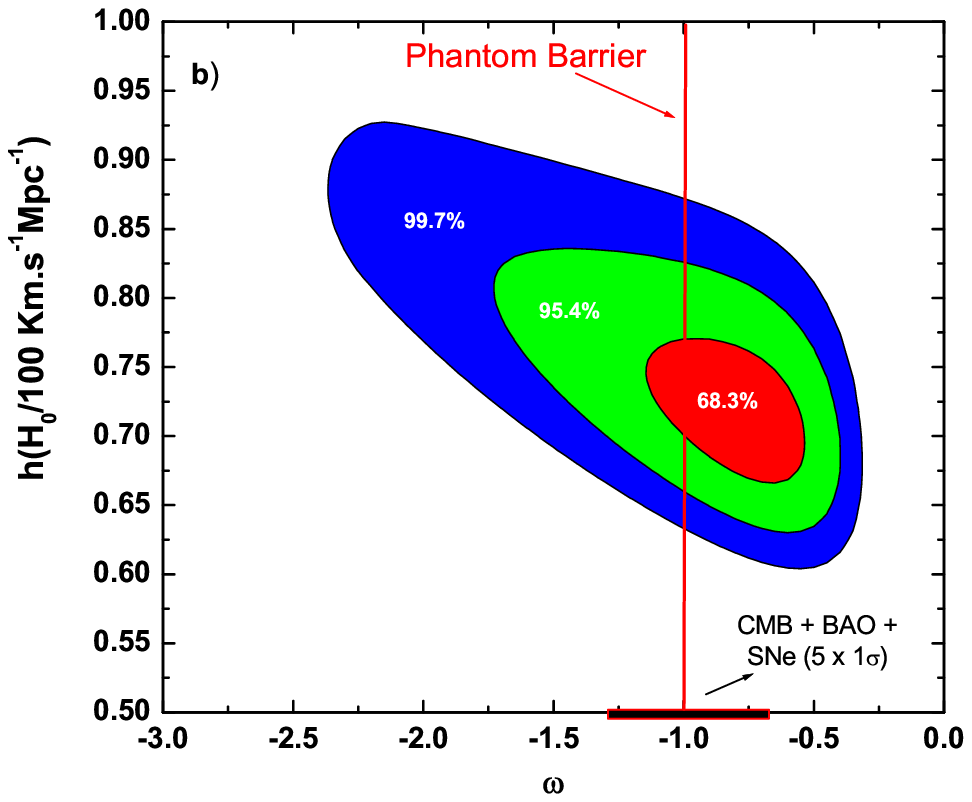,width=2.3truein,height=2.1truein}
\epsfig{figure=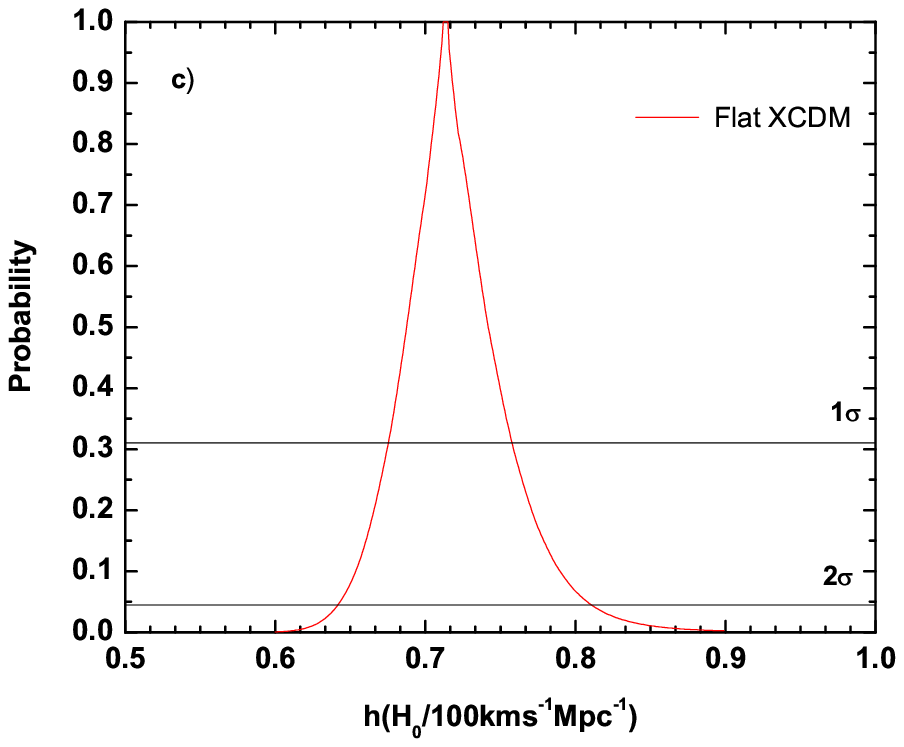,width=2.3truein,height=2.1truein}
 \hskip
0.1in} \caption{Flat XCDM Model. {\bf a)} The contour on the $(
h,\omega)$ plane (marginalized in $\Omega_M$), using 25 angular
diameter distances from De Fillips et al sample \cite{DeFilipp05}.
{\bf b)} Contours on the $(\omega, h)$ plane derived from a joint
analysis(SZE/X-Ray, BAO and Shift). Note that the values beyond the
phantom barrier  are also allowed (left of the dotted line). The
contours correspond to $68.3$\%, $95.4$\% and $99.7$\% c.l., and we
have marginalized over $\Omega_M$.  {\bf c)} Probability of the $h$
parameter. The horizontal cuts represent $68.3$\% and $95.4$\%. In
this case, we have marginalized over $\omega$ and $\Omega_M$.}
\end{figure*}

In  Figure 2b, we display the $(\omega, h)$ plane for the joint
analysis involving SZE/X-ray + BAO + Shift, and marginalizing over
the possible values of $\Omega_M$. In this figure, one may see the
observational limits to the phantom behavior of the dark energy in
our analysis for any $\Omega_M$ parameter. In spite of the phantom
dark energy to have a large region permitted for $2\sigma$ and
$3\sigma$, right of the phantom barrier with $\omega\geq -1$.
However, the best-fit value is left of the phantom barrier
$\omega=-0.76$ outside of the phantom energy zone. In accord with
recent results from WMAP plus others tests \cite{komatsu}.

In Figure 2c, we plot the likelihood function for the $h$ parameter
by marginalizing in $\omega$ and $\Omega_m$. The horizontal lines
are cuts of $68.3$\% and $95.4$\% probability. The agreement with
the results using flat and non-flat $\Lambda$CDM models, see figs.
(1c) (this paper) and (4) (previous work), is an evidence for the
robustness of the SZE/X-ray $h$ measurements.


\subsection{Comparing Results}

In the last few years, several measurements of $h=H_0/100$ were
presented in the literature, obtained using the SZE/X-ray method and
fixing the cosmology (using the cosmic concordance model): Carlstrom
{\it{et al.}} (2002) compiled distance determinations to 26 galaxy
clusters, and obtained $h = 0.60 \pm 0.03 km s^{-1}Mpc^{-1}$
\cite{carlstrom2002,usan04}; Mason {\it{et al.}} (2001), using 5
clusters, gives $h=0.66^{+0.14}_{-0.11}$ \cite{Mas01}; Reese and
coauthors (2002), using 18 clusters, found $h=0.60\pm 0.04$
\cite{Reese02}; Reese (2004), with 41 clusters, obtains $h\approx
0.61\pm 0.03$ \cite{Reese04}; Jones {\it{et al.}} (2005), using a
sample of 5 clusters, obtained $h=0.66^{+0.11}_{-0.10}$
\cite{Jones05}; and Schmidt {\it{et al.}} (2004) obtain a best-fit
$h = 0.69 \pm 0.08$ \cite{schmidt}.

The mean value of $h$ from the SZE/X-Ray measurements above appears
systematically lower than those estimated with other methods: e.g.
$h = 0.72 \pm 0.08$ from the Hubble Space Telescope (HST) Project
\cite{WFred01}, and $h = 0.73 \pm 0.03$ from the CMB anisotropy
\cite{komatsu}. On the other hand, the recent work of Sandage
{\it{et al.}} (2006) \cite{Sandage06}, from type Ia Supernovae,
predicts $h = 0.62 \pm 0.013 (random) \pm 0.05 (systematics)$. Yet,
we must to point out the work of Bonamente {\it{et al.}} (2006)
\cite{Boname06}, using 38 clusters and the SZE/X-ray method (with
the cosmic concordance model), which obtains
$h=0.769^{+0.039}_{-0.034}$. In this concern, we also recall that
Alcaniz \cite{Alc04} used 17 data between $0.14<z<0.78$ from
SZE/X-Ray angular distances to constrain the $\omega$ parameter.
Performing a joint analysis involving SZE/X-Ray + SNe Ia + CMB data
and fixing $\Omega_M=0.27$, he obtained
$\omega=-1.2^{+0.11}_{-0.18}$ at $68.3$\% c.l. ($1$ free parameter).
In a point of fact, our main interest here is to determine constraints
on the Hubble parameter independent of  SNe Ia data because
such observations already provide a very precise determination
of the Hubble parameter (see \cite{riess2009} and Refs. there in).

On the other hand, since the assumed cluster shape affects
considerably the SZE/X-ray distances, and, therefore, the $H_{0}$
estimates, we also compare our results with the ones of Holanda {\it
et al.} \cite{holanda07}.  These authors  used 38 angular diameter
distances from galaxy clusters, where a spherical $\beta$ model was
assumed to describe the clusters + BAO in a flat $\Lambda$CDM model.
They found $h=0.765 \pm 0.035$, a value slightly larger than the
ones found in this work.  We also  stress that the constraints on
the  Hubble parameter and $\Omega_M$ derived here using the
$\Lambda$CDM (free geometry) and the flat XCDM model, are in
agreement with the independent measurements provided by the WMAP
team \cite{komatsu}, the HST Project \cite{WFred01}, the work of
Bonamente {\it{et al.}} (2006) \cite{Boname06} and, more recently,
Riess {\it et al.} \cite{riess2009}.

In Table \ref{tables1}, we summarize the $H_0$ constraints based on
the SZE/X-ray technique. All estimates, except the last five lines,
are obtained in the framework of the flat $\Lambda$CDM model
($\Omega_m=0.3$, $\Omega_{\Lambda}=0.7$ - known as ``cosmic
concordance"). In the 8th and 9th lines, we see the results of CML
paper \cite{Cunha07}. They used a flat $\Lambda$CDM model and 24
galaxy clusters, with and without BAO (here labeled as ``Cunha
{\it{et al.}} 2007'' and ``Cunha {\it{et al.}} 2007 + BAO'',
respectively). In the 10th line, we display the results of Holanda
{\it et al.} using 38 galaxy clusters (spherical symmetry) + BAO in
a flat $\Lambda$CDM \cite{holanda07}. In the last two lines we
present our results derived through a joint analysis involving
SZE/X-ray,  BAO and CMB shift parameter.

\begin{table}[htbp]
\caption{Constraints on the ``little" Hubble parameter  $h$  based
on  SZE/X-ray technique applied to galaxy clusters data.}
\label{tables1}
\par
\begin{center}
\begin{tabular}{|c||c|c|c|}
\hline\hline Reference & $\Omega_m$ & $h$ ($1\sigma$) &
$\chi_{red.}^2$
\\ \hline\hline Mason \textit{et al.} 2001 & $0.3$ &
$0.66^{+0.14}_{-0.11}$ & $0.35$
\\ Carlstrom {\it{et al.}} 2002 & $0.3$ &
$0.60^{+0.03}_{-0.03}$ & --
\\ Reese \textit{et al.} 2002 & $0.3$ &
$0.60^{+0.04}_{-0.04}$
& $0.97$ \\
Reese 2004 & $0.3$ & $0.61^{+0.03}_{-0.03}$ & -- \\
Jones \textit{et al.} 2005 & $0.3$ & $0.66^{+0.11}_{-0.10}$
& -- \\
Schmidt {\it{et al.}} 2004 & $0.3$ & $0.69^{+0.08}_{-0.08}$
& -- \\
Bonamente \textit{et al.} 2006 & $0.3$ & $%
0.77^{+0.04}_{-0.03}$ & $0.83$ \\
Cunha \textit{et al.} 2007 & $0.15^{+0.57}_{-0.15}$ & $0.75^{+0.07}_{-0.07}$ & $1.06$ \\
Cunha \textit{et al.} 2007 + BAO & $0.27^{+0.04}_{-0.03}$ & $0.74^{+0.04}_{-0.03}$ & $1.06$ \\
Holanda \textit{et al.} 2007 + BAO& $0.27^{+0.04}_{-0.03}$ & $0.765^{+0.035}_{-0.035}$ & $0.96$ \\
\textbf{This Work ($\Lambda$CDM )} & {\boldmath{$0.272^{+0.03}_{-0.02}$}} & {\boldmath%
{$0.732^{+0.042}_{-0.037}$}} & {\boldmath{$1.12$}} \\
\textbf{This Work (flat-XCDM)} & {\boldmath{$0.30^{+0.05}_{-0.04}$}} & {\boldmath%
{$0.714^{+0.044}_{-0.034}$}} & {\boldmath{$1.13$}} \\
\hline\hline
\end{tabular}
\end{center}
\end{table}

\section{Conclusions}

In this work we have discussed a determination of the Hubble
 parameter and other relevant cosmological quantities  based on the
 SZE/X-ray distance technique for a sample of
25 clusters compiled by De Filippis {\it{et al.}} \cite{DeFilipp05}.
In order to prove the robustness of  the $H_0$ parameter we relaxed
the flatness condition of the $\Lambda$CDM cosmology. The degeneracy
on the cosmological parameters was broken using BAO and shift
parameter. While the former test is independent of $H_0$ the last
one is weakly dependent. By comparing the results of this work with
the ones of Cunha \textit{et al.} \cite{Cunha07} (see Table
\ref{tables1}), which uses only a flat $\Lambda$CDM model and BAO,
we clearly see that the present estimates of  $H_0$ are virtually
independent of the geometry of the Universe. For a general
$\Lambda$CDM cosmology we obtain $H_0=73.2^{+4.3}_{-3.7}$ km
s$^{-1}$ Mpc$^{-1}$(1$\sigma$ - without systematic errors)

In order to test the real dependence of the method with the adopted
cosmology, we have compared the constraints from $\Lambda$CDM model
with the flat XCDM. In the same way, we also conclude that the $H_0$
estimates presents a negligible dependence on dark energy models
with constant $\omega$. For a flat XCDM we obtain
$H_0=71.4^{+4.4}_{-3.4}$ km s$^{-1}$ Mpc$^{-1}$(1$\sigma$ - without
systematic errors). We study also possible observational limits to
the phantom behavior of the dark energy for these SZE/X-Ray data
plus BAO and Shift Parameter. We results indicate a large region
permitted by phantom energy ($\omega\leq -1$), but the best-fit
value is $\omega=-0.76$ outside of this phantom energy zone.

The constraints on the Hubble parameter derived here  are also
consistent with some recent cosmological observations like the WMAP
and the HST Key Project. Implicitly, such an agreement suggests that
the elliptical morphology describing the cluster sample and the
associated isothermal $\beta$-model is quite realistic.

Finally, we stress that the combination of these four independent
phenomena (SZE, X-Ray, BAO and Shift) provides an interesting method
to constrain the Hubble constant, and more important, it is
independent of any calibrator usually adopted in the determinations
of the distance scale.

\section*{Acknowledgments}
The authors are grateful to J. F. Jesus and V. C. Busti for helpful
discussions. RFH and JVC are supported by FAPESP under grants 07/52912-2 and 05/02809-5,
respectively, and JASL is partially supported by CNPq (No. 304792/2003)
and FAPESP (No. 04/13668).


\begin{thebibliography}{99}

\bibitem{SunZel72} R. A. Sunyaev and Ya. b. Zel'dovich, Comments Astrophys. Space Phys. {\bf 4},
173 (1972).
\bibitem{sunzel80} R. A. Sunyaev and Ya. b. Zel'dovich, MNRAS, {\bf190}, 413 (1980).

\bibitem{itoh} N. Itoh, Y. Kohyama and S. Nozawa, Astrophys. J.,  {\bf 505}, 7 (1998).

\bibitem{Bartlett1} Bartlett J. G., Silk J., Astrophys. J. {\bf 423}, 12
(1994).
\bibitem{Bartlett2} J.G. Bartlett J. G., Astrophys. Space Sci. {\bf 290},
105 (2004).

\bibitem{Kohyama09} Y. Kohyama and S. Nozawa, Phys. Rev. D {\bf 79}, 083005
(2009).
\bibitem{Lavalle10} J. Lavalle, C. Boehm, J. Barthes JCAP {\bf 1002}, 005 (2010).

\bibitem{Birki99} M. Birkinshaw, Phys. Rep. {\bf 310}, 97 (1999).

\bibitem{Carlstrom02} J. E. Carlstrom, G. P. Holder, and E. D. Reese,
Ann. Rev. Astron. and Astrophys. {\bf 40}, 643 (2002).

\bibitem{Jones05} M. E. Jones et al., Mon. Not. Roy. Astro. Soc. {\bf 357}, 518 (2005).

\bibitem{Freedman00} W. L. Freedman, Phys. Rep. \textbf{333}, 13
(2000);  J. A. Peacock, {\it{Cosmological Physics}} (Cambridge Univ.
Press, Cambridge, 1999).

\bibitem{komatsu} Komatsu, E. {\it  et al.} 2010, astro-ph.CO:1001.4538 (WMAP
collaboration).

\bibitem{Hu05} W. Hu, in {\it{ASP Conf. Ser. 339:
Observing Dark Energy}}, edited by S. C. Wolf \& T. R. Lauer, p. 215
(2005); M. Tegmark {\it{et al.}}, Phys. Rev. D \textbf{69}, 103501
(2004).

\bibitem{Sandage06} A. Sandage {\it{et al.}}, Astrophys. J. \textbf{653},
843 (2006); G. A. Tammann {\it{et al.}}, ApJ \textbf{679}, 52
(2008).

\bibitem{VanLeeuwen07} F. van Leeuwen {\it{et al.}}, MNRAS \textbf{379}, 723 (2007).

\bibitem{WFred01} W. L. Freedman {\it{et al.}}, Astrophys. J. \textbf{553}, 47 (2001).

\bibitem{riess2009} A. G. Riess {\it{et al.}}, Astrophys. J. \textbf{699}, 534 (2009).

\bibitem{FalelySuyu10} R. Fadely {\it et al.}, Astrophys. J. {\bf{711}}, 246 (2010).

\bibitem{suyu} S. H. Suyu {\it et al.}, ApJ {\bf{711}}, 201 (2010).

\bibitem {Jackson07} N. Jackson, Liv. Rev. Rel. {\bf 10}, 4 (2007).

\bibitem{Cunha07} J. V. Cunha, L. Marassi and J. A. S. Lima, Mon. Not. Roy. Astro. Soc. \textbf{379}, L1-L5 (2007), [astro-ph/0611934].


\bibitem{DeFilipp05} E. De Filippis, M. Sereno, M. W. Bautz and G. Longo, Astrophys. J. \textbf{625},
108 (2005).

\bibitem{Eisenstein05} D. J. Eisenstein {\it{et al.}}, Astrophys. J. \textbf{633},
560 (2005).

\bibitem{Percival10} W. Percival et al.,
Mon. Not. Roy. Astron. Soc., {\bf 401}, 2148 (2010).


\bibitem{Kobayashi96} S. Kobayashi, S. Sasaki and Y. Suto, Publ. Astron. Soc. Jap. {\bf{48}}, L107 (1996).

\bibitem{carlstrom2002} J. E. Carlstrom, G. P. Holder and E. D. Reese, ARA\&A \textbf{40},
643 (2002).

\bibitem{Mas01} B. S. Mason {\it{et al.}}, ApJ \textbf{555}, L11 (2001).

\bibitem{Reese02} E. D. Reese {\it{et al.}}, ApJ \textbf{581}, 53 (2002).

\bibitem{Reese04} E. D. Reese, in {\it{Measuring and Modeling the Universe}}, ed. W. L. Freedman (CUP) p. 138 (2004),
[astro-ph/0306073].

\bibitem{schmidt} R. W. Schmidt, S. W. Allen and A. C. Fabian,  Mon. Not. Roy. Astron. Soc.  \textbf{352},
1413 (2004).

\bibitem{Boname06} M. Bonamente {\it{et al.}}, Astrophys. J. \textbf{647}, 25 (2006).

\bibitem{Itoh9806} S. Nozawa, N. Itoh, Y. Suda and Y. Ohhata, Nuovo Cimento B
\textbf{121}, 487 (2006); N. Itoh, Y. Kohyama and S. Nozawa,
Astrophys. J. \textbf{502}, 7 (1998); S. Nozawa, N. Itoh    and Y.
Kohyama, Astrophys. J. \textbf{508}, 17 (1998).

\bibitem{perm98} S. Perlmutter {\it et al.}, Nature {\bf{ 391}}, 51 (1998);
 S. Perlmutter {\it et al.}, Astrophys.  J. {\bf{517}}, 565 (1999).

\bibitem{Riess} A. G. Riess {\it et al.}, Astron. J. {\bf{116}}, 1009 (1999).

\bibitem{Astier06}  P. Astier {\it et al.}, Astron. Astrophys.
 {\bf 447}, 31 (2006); A. G. Riess {\it et al.}, Astrop. J. {\bf 659}, 98 (2007).

\bibitem{bar10} M. Bartelmann, Rev. Mod. Phys. {\bf 82}, 331 (2010).

\bibitem{Lambdat} W. Chen and Y-S. Wu, Phys. Rev. D {\bf 41}, 695 (1990); D. Pav\'{o}n, Phys.
Rev. D {\bf 43}, 375 (1991); J. C. Carvalho, J. A. S. Lima and I.
Waga, Phys. Rev. D {\bf{46}}, 2404 (1992); J. A. S. Lima and J. M.
F. Maia, Phys. Rev D {\bf 49}, 5597 (1994);  J. A. S. Lima and J. M.
F. Maia, Phys. Rev. D {\bf 65}, 083513 (2002), [astro-ph/0112091];

\bibitem{xmatt} T. Padmanabhan and T. R. Choudhury,
Mon. Not. Roy. Astron. Soc. {\bf{344}}, 823 (2003); P. T. Silva and
O. Bertolami, Astrophys.  J.  {\bf{599}}, 829 (2003); J. A. S. Lima,
J. V. Cunha and J. S. Alcaniz, Phys. Rev. D {\bf 68}, 023510 (2003),
[astro-ph/0303388];  Z.-H. Zhu and M.-K. Fujimoto, \apj {\bf{585}},
52 (2003); R. Colistete Jr and J. C. Fabris,
    Class. Quant. Grav. {\bf {22}}, 2813 (2005); B. Feng, X. Wang and X.  Zhang, Phys. Lett. B {\bf{607}},
     35 (2005).

\bibitem{scalarf}  C. Li, D. E. Holz, A. Cooray, Phys. Rev. D \textbf{75}, 103503
(2007); X. Zhang, Phys. Rev. D \textbf{74}, 103505 (2006); H.
Ziaeepour, AIP Conf. Proc. \textbf{861}, 1059 (2006),
[hep-ph/0604014]; F. C. Carvalho {\it et al.}, Phys. Rev. Lett. {\bf
97},  081301 (2006), [astro-ph/0608439]; [arXiv:0704.3043]; L. R.
Abramo and N. Pinto-Neto, Phys. Rev. D {\bf 73}, 063522 (2006).


\bibitem{Chap1}
M. Makler, S. Q. de Oliveira, and I. Waga, Phys. Lett B., {\bf 555},
1, (2003); M. C. Bento, O. Bertolami, and A. A. Sen, Phys. Lett. B.,
{\bf 575}, 172, (2003); A. Dev, J. S. Alcaniz, and D. Jain, Phys.
Rev. D., {\bf 67}, 023515, (2003); Y. Gong, C. K. Duan, Mon. Not.
Roy. Astron. Soc., {\bf 352}, 847, (2004); G. Sethi {\it{et al.}},
IJMP D \textbf{15}, 1089 (2006);  A. A. Sen, R. J. Scherrer,
Phys.Rev. D \textbf{72}, 063511 (2005).

\bibitem{Chap2}Z. H. Zhu, Astron. Astrophys., {\bf 423}, 421, (2004). L.
Amendola, I. Waga, and F. Finelli, {\tt astro-ph/0509099}; J. A. S.
Lima, J. S. Alcaniz and J. V. Cunha, Astrop.  Phys. {\bf 31}, 233
(2009).

\bibitem{CCDM} G. Steigman, R. C. Santos and J.A.S. Lima, JCAP {\bf 06}, 033
(2009), arXiv:0812.3912 [astro-ph]; J. A. S. Lima, J. F. Jesus and
F. A. Oliveira, {\tt arXiv:0911.5727}, (2009); S. Basilakos, and J.
A. S. Lima,  arXiv:1003.5754 [astro-ph.CO], (2010);  S. Debnath, A.
K. Sanyal, arXiv:1005.3933 [astro-ph.CO], (2010).


\bibitem{Efsta99} G. Efstathiou, Mon. Not. Roy. Astron. Soc. \textbf{310}, 842 (1999);
J. V. Cunha, L. Marassi and  R. C. Santos, IJMP D {\bf 16}, 403
(2007), [astro-ph/0306319].

\bibitem{Phantom} V. Faraoni, Int. J. Mod. Phys. D 11, 471 (2002); R. R.
Caldwell, Phys. Lett. B \textbf{545}, 23 (2002); R. R. Caldwell, M.
Kamionkowski, and N. N. Weinberg, Phys. Rev. Lett. 91, 071301
(2003); J. A. S. Lima and J. S. Alcaniz, Phys. Lett. B {\bf 600},
191 (2004), astro-ph/0402265;   G.~Izquierdo and D.~Pavon, Phys.\
Lett.\  B {\bf 639}, 1 (2006), [arXiv:gr-qc/0606014];  S. H. Pereira
and J. A. S. Lima, Phys. Lett. B {\bf 669}, 266 (2008),
arXiv:0806.0682 [astro-ph].

\bibitem{Bond:1997wr}  J. R. Bond, G. Efstathiou and M. Tegmark,
  Mon.\ Not.\ Roy. Astron. Soc.\  {\bf 291}, L33 (1997).

\bibitem{Nesseris:2006er}
  S. Nesseris and L. Perivolaropoulos,
  JCAP {\bf 0701}, 018, (2007).

\bibitem{Davis} T. M. Davis {\it{et al.}}, ApJ \textbf{666}, 716D (2007), [astro-ph/0701510].

\bibitem{Elgaroy} O. Elgar{\o}y and T. Multam{\"a}ki, A\&A \textbf{471}, 65E (2007),
[astro-ph/0702343].

\bibitem{Lima03} J. A. S. Lima, J. V. Cunha and J. S. Alcaniz, Phys. Rev.
D \textbf{68}, 023510 (2003).

\bibitem{LA00} J. A. S. Lima and J. S. Alcaniz, Astron. Astrophys. {\bf
357}, 393 (2000), astro-ph/0003189;  {\bf ibdem}, Astrophys.\ J.\
{\bf 566}, 15 (2002), [arXiv:astro-ph/0109047];

\bibitem{Ebeling96} H. Ebeling {\it{et al.}},  Mon. Not. Roy. Astron. Soc.  \textbf{281}, 799 (1996).

\bibitem{sze1} Y. Wang and P. Mukherjee, Phys. Rev. D \textbf{76}, 103533 (2007).

\bibitem{usan04} J. P. Uzan, N. Aghanim, and Y. Mellier, Phys. Rev. D \textbf{70},
083533 (2004).

\bibitem{Alc04} J. S. Alcaniz, Phys. Rev. D{\bf 69}, 083521 (2004).

\bibitem{holanda07} R. F. L. Holanda, J. V. Cunha and J. A. S. Lima, 2008 [astro-ph/0807.0647].

\end{thebibliography}
\end{document}